\documentclass[10pt,conference]{IEEEtran}
\usepackage{epsfig,setspace,amsmath,epsf,amssymb,bm,theorem,cite,graphicx, epstopdf,algorithm,algpseudocode,float,color,mathtools,authblk,tikz,physics}
\usetikzlibrary{positioning}
\usetikzlibrary{decorations.text}
\usetikzlibrary{decorations.pathmorphing}
\usepackage{booktabs}
\usepackage{arydshln}
\usepackage[short,c2]{optidef}

\newtheorem{remark}{Remark}
\newtheorem{lemma}{Lemma}

\newenvironment{Proof}[1]{\medskip\par\noindent{\bf Proof:\,}\,#1}{{\mbox{\,$\blacksquare$}\par}}

\IEEEoverridecommandlockouts
\allowdisplaybreaks

\begin{document}

\title{AoII-Optimum Sampling of CTMC Information Sources Under Sampling Rate Constraints}

\author[1]{Ismail Cosandal}
\author[2]{Nail Akar}
\author[1]{Sennur Ulukus}

\affil[1]{\normalsize University of Maryland, College Park, MD, USA}
\affil[2]{\normalsize Bilkent University, Ankara, T\"{u}rkiye}

\maketitle

\let\thefootnote\relax\footnotetext{This work is done when N.~Akar is on sabbatical leave as a visiting professor at University of Maryland, MD, USA, which is supported in part by the Scientific and Technological Research Council of T\"{u}rkiye  (T\"{u}bitak) 2219-International Postdoctoral Research Fellowship Program.}

\begin{abstract}
We consider a sensor that samples an $N$-state continuous-time Markov chain (CTMC)-based information source process, and  transmits the observed state of the source, to a remote monitor tasked with timely tracking of the source process. The mismatch between the source and monitor processes is quantified by age of incorrect information (AoII), which penalizes the mismatch as it stays longer, and our objective is to minimize the average AoII under an average sampling rate constraint. We assume a perfect reverse channel and hence the sensor has information of the estimate while initiating a  transmission or preempting an ongoing transmission. First, by modeling the problem as an average cost constrained semi-Markov decision process (CSMDP), we show that the structure of the problem gives rise to an optimum threshold policy for which the sensor initiates a transmission once the AoII exceeds a threshold depending on the instantaneous values of both the source and monitor processes. However, due to the high complexity of obtaining the optimum policy in this general setting, we consider a relaxed problem where the thresholds are allowed to be dependent only on the estimate. We show that this relaxed problem can be solved with a novel CSMDP formulation based on the theory of absorbing MCs, with a computational complexity of $\mathcal{O}(N^4)$, allowing one to obtain optimum policies for general CTMCs with over a hundred states.
\end{abstract}

\section{Introduction} \label{sec:introduction}
We consider a source equipped with a sensor sampling an $N$-state ($2 \leq N < \infty$) continuous-time Markov chain (CTMC)-based information source process, which transmits the observed state of the source to a remote monitor tasked with tracking the source process. For such remote estimation problems, age of incorrect information (AoII) was first proposed in \cite{maatouk2020} as a mismatch and information freshness metric allowing an increase in dissatisfaction with respect to time as long as the source and monitor processes stay out of sync. In this paper, we use AoII in a more limited scope (than the more general setting of \cite{maatouk2020}) where dissatisfaction changes linearly with time. In this regard, AoII is different from conventional mismatch metrics such as mean square error (MSE) or binary freshness \cite{bastopcu2021,cosandal2023timely} for which there is no time-dependent dissatisfaction. AoII is also fundamentally different from other well-established information freshness metrics including age of information (AoI) and related metrics such as version age of information \cite{bastopcu2020should}, value of information \cite{ayan2019value}, etc.,  since the AoII process can be reset when the out of sync source and monitor processes turn out to be in sync back again, not by means of a status update only, but also by the source process making a state transition on its own, leading to in sync status.  

The AoII metric is mostly investigated in a discrete-time setting. In \cite{maatouk2020,kam2020age,maatouk2022age,chen2022preempting,chen2021minimizing}, system models with discrete-time Markov chain (DTMC) sources are studied. In these works, optimum transmission policies are obtained to minimize the average AoII under sampling rate constraints, in the form of a threshold policy, by using Markov decision processes (MDP). However, only special DTMCs are studied in these works. In particular, references \cite{maatouk2020} and \cite{kam2020age} study binary sources, whereas in \cite{maatouk2022age} and \cite{chen2022preempting}, DTMCs have general number of states, but they are assumed to be symmetric. Reference \cite{chen2021minimizing} studies a source process modeled as a birth-death Markov process. On the other hand, in \cite{saha2022relationship}, the information source is piece-wise linear and the relation between AoII and MSE is investigated. In another related work \cite{joshi2021minimization}, AoII is used for an auto-regressive Markov process, and a new metric called age of incorrect estimation (AoIE) is proposed. There have been fewer studies considering the continuous-time setting. In \cite{inoue2019aoi}, the probability distribution of AoI, and the joint probability distribution of the observed process and its estimation are derived. Pull-based sampling of multiple heterogeneous CTMC-based sources are considered  in \cite{akar2023optimum} where the monitors query the sources according to a Poisson process and optimum sampling rates for each source are obtained that maximize several information freshness metrics including weighted binary freshness. In our previous work \cite{cosandal2024modeling}, we proposed an analytical model for push-based (and also pull-based) sampling systems in which the sensor waits for the AoII to reach a random threshold before transmitting an update and it was shown in \cite{cosandal2024modeling} that average AoII performance improves when the variance of this threshold decreases. In this work, we extend the analytical model of \cite{cosandal2024modeling} to deterministic thresholds and we additionaly obtain the optimum thresholds.

The focus of this paper is a sensor observing a finite state-space CTMC source process $X(t) \in \mathcal{N} = \{1,2,\ldots,N\}$, and a monitor process $\hat{X}(t)$ with the same state-space, with status updates transmitted via a communications channel characterized with exponentially distributed service times, which is illustrated in Fig.~\ref{fig:SystemModel}. The dynamics of the CTMC is known to the source in which case our goal is to develop a sampling policy in order to minimize the average AoII under an average sampling rate constraint.
First, we investigate the optimum transmission and preemption policy by using a constrained semi-Markov decision process (CSMDP) framework with the Lagrangian approach under the assumption that such decisions can be made at any time \cite{white1993mdp,ibe2013markov,hu2007markov}. Our structural analysis shows that the optimum transmission policy is a threshold policy for which the sensor initiates a transmission once the AoII process exceeds a threshold depending on the instantaneous values of both the source and monitor processes $X(t)$ and $\hat{X}(t)$. This policy reduces to a single threshold policy for a symmetric process, and a threshold policy only depending on $\hat{X}(t)$, for binary sources. Since the solution is highly complex for the most general scenario involving general $N$-state sources with non-symmetric states, we focus our attention to a relaxed update system where these thresholds are allowed to depend only on $\hat{X}(t)$. This relaxed problem is formulated as a CSMDP based on a novel construction of an SMDP making state transitions at the embedded epochs of synchronization points, using the theory of absorbing CTMCs. Employing the Lagrangian approach in \cite{makowski1986} for the underlying CSMDP, we obtain optimum transmission policies while minimizing the average AoII under average sampling rate constraints.

Our contributions can be summarized as follows. 1) While similar system models in discrete-time are studied for only special DTMCs, e.g., binary sources, symmetric states, we study in this paper general $N$-state CTMC-based information sources. 2) For the relaxed problem where the thresholds are allowed to depend only on $\hat{X}(t)$, we propose a novel CSMDP formulation with continuous action space by means of constructing a semi-Markov process (for any given vector of thresholds) making state transitions at the embedded epochs of synchronization points.
3) Using the well-established policy iterations and the particular use of bisection search for policy improvement to cope with continuous action space, and the Lagrangian approach, we propose a method to solve the  CSMDP to obtain the optimum threshold values with a reasonable complexity of $\mathcal{O}(N^4)$ allowing us to obtain optimum policies for moderate-sized CTMC sources.

\begin{figure}[tb]
	\centering
	\begin{tikzpicture}[scale=0.29]
\draw[rounded corners,thick,darkgray,dashed] (0,-1) rectangle (7,4) {};
\filldraw (3.5,2) node[anchor=center] {\small{source process}};
\filldraw (3.5,0.75) node[anchor=center] {\small{$X(t)$}};
\draw[rounded corners,thick,darkgray,dashed] (21.5,-1) rectangle (29,4) {};
\filldraw (25.25,2) node[anchor=center] {\small{monitor process}};
\filldraw (25.25,0.75) node[anchor=center] {\small{$\hat{X}(t)$}};
\draw[ultra thick,->,black] (21.5,-0.45) -- (7,-0.45);
\filldraw (14.5,-0.25) node[anchor=north] {\small{instantaneous feedback}};
\draw[ultra thick] (7,2) -- (9,2);
\draw[ultra thick] (9,2) -- (11,4);
\draw[->,blue,ultra thick] (11,4) -- (13,2);
\node at (11,2) (nodeA) {};
\node at (16,7) (nodeB) {};
\draw[ultra thick,->] (13,2) -- (15.25,2);
\draw [decoration={text along path,
    text={|\scriptsize|transmit},text align={center}},decorate]  (nodeA) -- (nodeB);

\draw[->,red,ultra thick,dashed] (12.5,1.5) -- (10.5,3.5);
\node at (8,-1) (nodeA) {};
\node at (12,3) (nodeB) {};

\draw [decoration={text along path,
    text={|\scriptsize|preempt},text align={center}},decorate]  (nodeA) -- (nodeB);

\draw[rounded corners,thick,darkgray] (15.25,0.5) rectangle (19.25,3.5) {};
\filldraw (17.25,2.5) node[anchor=center] {\small{channel}};
\filldraw (17.25,1.5) node[anchor=center] {$\mu$};
\draw[ultra thick,->] (19.25,2) -- (21.5,2);

	\end{tikzpicture}
	\caption{A remote estimation system with the source process $X(t)$ and the monitor process $\hat{X}(t)$ for which the source employs a transmission policy to transmit the status update packets via the channel, and also a preemption policy to preempt ongoing transmissions when the observed information becomes obsolete.}
	\label{fig:SystemModel}
\end{figure}
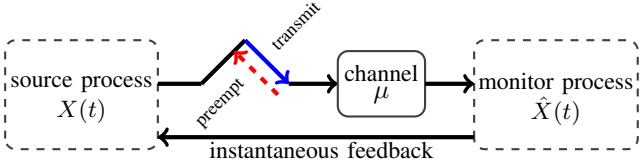

\begin{figure}[tb]
	\centering
	\begin{tikzpicture}[scale=0.24]
\draw[->] (3,8) -- (27,8);
\filldraw (27,8) node[anchor=north] {\small{$t$}};
\draw[->] (3,8) -- (3,17);

\draw[-] (5,8) -- (5,7.65);
\draw[-] (7,8) -- (7,7.65);
\draw[-] (9,8) -- (9,7.65);
\draw[-] (11,8) -- (11,7.65);
\draw[-] (13,8) -- (13,7.65);
\draw[-] (15,8) -- (15,7.65);
\draw[-] (17,8) -- (17,7.65);
\draw[-] (19,8) -- (19,7.65);
\draw[-] (21,8) -- (21,7.65);
\draw[-] (23,8) -- (23,7.65);
\draw[-] (25,8) -- (25,7.65);

\filldraw (7,8) node[anchor=north] {\scriptsize{2}};
\filldraw (11,8) node[anchor=north] {\scriptsize{4}};
\filldraw (15,8) node[anchor=north] {\scriptsize{6}};
\filldraw (19,8) node[anchor=north] {\scriptsize{8}};
\filldraw (23,8) node[anchor=north] {\scriptsize{10}};
\filldraw (25,8) node[anchor=north] {\scriptsize{11}};
\filldraw (5,8) node[anchor=north] {\scriptsize{1}};
\filldraw (9,8) node[anchor=north] {\scriptsize{3}};
\filldraw (13,8) node[anchor=north] {\scriptsize{5}};
\filldraw (17,8) node[anchor=north] {\scriptsize{7}};
\filldraw (21,8) node[anchor=north] {\scriptsize{9}};

\draw[red, ultra thick] (3,18) -- (6,18);
\filldraw (6,18) node[anchor=west] {\small{AoII$(t)$}};
\draw[lightgray, ultra thick] (12,18) -- (15,18);
\filldraw (15,18) node[anchor=west] {\small{$X(t)$}};
\draw[dotted, red, very thick] (21,18) -- (24,18);
\filldraw (24,18) node[anchor=west] {\small{$\hat{X}(t)$}};

\draw[->,blue,very thick] (7,14) -- (7,12);
\draw[->,blue,very thick] (21,14) -- (21,16);

\filldraw (3,12) node[anchor=east] {\small{1}};
\draw[lightgray,ultra thick] (3,12) -- (11,12);
\filldraw (3,16) node[anchor=east] {\small{2}};
\draw[dotted,red,very thick] (3,16) -- (7,16);

\draw[lightgray,ultra thick] (11,16) -- (17,16);
\draw[lightgray,ultra thick] (17,12) -- (19,12);
\draw[lightgray,ultra thick] (19,16) -- (25,16);
\draw[dotted,red,very thick] (7,12) -- (21,12);
\draw[dotted,red,very thick] (21,16) -- (25,16);

\draw[red,ultra thick] (3,8) -- (7,12);
\draw[red,ultra thick] (7,12) -- (7,8);
\draw[red,ultra thick] (7,8) -- (11,8);
\draw[red,ultra thick] (11,8) -- (17,14);
\draw[red,ultra thick] (17,14) -- (17,8);
\draw[red,ultra thick] (17,8) -- (19,8);
\draw[red,ultra thick] (19,8) -- (21,10);
\draw[red,ultra thick] (21,10) -- (21,8);
\draw[red,ultra thick] (21,8) -- (25,8);
\end{tikzpicture}
	\caption{A sample path of $X(t)$, $\hat{X}(t)$, and $\text{AoII}(t)$ (thick red solid curve) for an example scenario when $N=2$ and $\text{AoII}(0)=0$.
 The arrows represent the reception epochs of status update packets at the monitor.  Notice that $\text{AoII}(t)$ drops to zero at  $t=7$ without a reception.}
	\label{fig:SamplePath}
\end{figure}
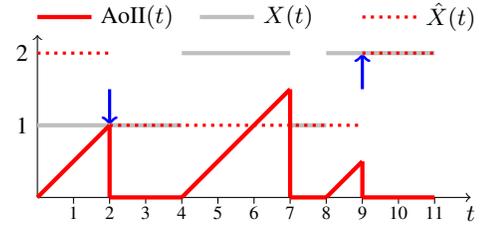

\section{System Model} \label{sec:2}
We consider a finite state-space CTMC process $X(t) \in \mathcal{N}=\{1,2,\ldots,N\}$ with generator $\bm Q = \{ q_{ij} \}$. $X(t)$ visits a given state $i \in \mathcal{N}$ for a duration denoted by $H_i$ which is exponentially distributed with parameter $\sigma_i = -q_{ii}$, i.e., $H_i \sim \text{Exp}(\sigma_i)$, after which a state transition takes place to state $j \in \mathcal{N}$ with transition probability  $p_{ij}=-{q_{ij}}/{q_{ii}}$, for $i\neq j$. As illustrated in Fig.~\ref{fig:SystemModel}, the sensor observes the source process $X(t)$ and it is allowed at any time to initiate a transmission of the sample value of $X(t)$ via a communications channel which is abstracted by an exponentially distributed service time distribution with parameter $\mu$. The estimator law is $\hat{X}(t) = X(t')$ where $t'$ is the timestamp of the latest status update packet received before time $t$, which is also widely used in existing remote estimation studies \cite{maatouk2020}. We assume instantaneous feedback from the monitor towards the source. Therefore, the source has perfect knowledge about $\hat{X}(t)$ and $\text{AoII}(t)$, and makes decisions accordingly. The only restriction for the source is that it should only transmit a value it is currently observing. Therefore, in our system model, the source preempts the ongoing transmission if the observed state is to change during transmission. The mismatch between $X(t)$ and $\hat{X}(t)$ is measured by using the AoII process with linear penalty as,
\begin{align}
\text{AoII}(t)=t-\max(\nu \; | \; X(\nu)=\hat{X}(\nu),\; \nu \leq t). \label{eq:aoii}
\end{align} 
A sample path of the processes $X(t)$, $\hat{X}(t)$, and $\text{AoII}(t)$ is illustrated for an example scenario
in Fig.~\ref{fig:SamplePath}. The average AoII, denoted by $\text{MAoII}$ is the time average, i.e., $\text{MAoII}= \frac{1}{T} \int_{t=0}^T \text{AoII}(t) \dd{t}$. Also let $N(t)$ denote the number of samples taken in the interval $[0,t]$. We define the average sampling rate $R$ as $R=\frac{1}{T} \int_{t=0}^T N(t) \dd{t}$. For the optimization problem studied in this paper, a budget is allocated for the average sampling rate, denoted by $b$. The optimization problem of interest is then stated as  
\begin{mini}
	{\phi}{\text{MAoII}^{\phi}} 
	{\label{Opt1}}
    {}
	\addConstraint{ R^{\phi} }{\leq b} 
\end{mini}
over all sampling policies $\phi$, with $\text{MAoII}^{\phi}$ (resp. $R^{\phi}$) denoting the average AoII (resp. sampling rate) obtained by imposing policy $\phi$.

We consider the source taking actions on infinitesimal time intervals of length $\delta$ if $X(t)$ and $\hat{X}(t)$ are the same during this time interval, or whenever $X(t)$ or $\hat{X}(t)$ changes. The $n$th decision epoch is denoted by $t_n$, and the time interval $t_{n+1}-t_n$ is a random variable called \emph{sojourn times}. We study the optimization problem in \eqref{Opt1} by modeling it as a semi-MDP (SMDP) which allows us to find the optimum policy when the sojourn times after each decision is a random variable. Then, we can define the state at the $n$th decision epoch as $s_n=(X(t_n),\hat{X}(t_n),\text{AoII}(t_n),Z(t_n))$, with state space  $\mathcal{S}=\{\mathcal{N}\times\mathcal{N}\times {\rm I\!R}_+ \times\{0,1\}\}$. While other terms are self-explanatory, $Z(t_n)$ indicates whether the transmission has already started, i.e., $Z(t_n)=1$, or not, i.e., $Z(t_n)=0$. There are three possible actions $u(t_n)$ in the action space $\mathcal{U}=\{-1,0,1\}$. While $u(t_n)=1$ and $u(t_n)=-1$ correspond to initiation of a transmission and preemption of an ongoing transmission, respectively, $u(t_n)=0$ means stay idle for $Z(t_n)=0$ or continue transmission for $Z(t_n)=1$. Notice that $u(t_n)=1$ (resp. $u(t_n)=-1$) is only feasible action when $Z(t_n)=0$ (resp. $Z(t_n)=1$) and immediately changes it to $Z(t_n)=1$ (resp. $Z(t_n)=0$). 

In this problem, between decision epochs, we define two different costs namely \emph{age cost}, and \emph{sampling cost}. Age cost corresponds to the area under the AoII curve until the next decision epoch, and sampling cost is the total number of initiated transmissions in the same time interval. For state $s$ and action $u$, they are denoted by $A_s(u)$ and $C_s(u)$, respectively. The value of $A_s(u)$ can be calculated for each given state and action with \eqref{eq:aoii} and this calculation is detailed more in Appendix~B, and the value of $C_s(u)$ equals to $1$ only when $u=1$. Finally, we can finalize the formulation by representing the transition probabilities from state $s$ to $s'$ for action $u$, with $p_{ss'}(u)$, and the sojourn time for state $s$ and action $u$ with $D_s(u)$. The optimization problem in \eqref{Opt1} can be solved by the Lagrangian method described in Appendix~A. Our main result is given in the following lemma.
\begin{lemma} \label{lem:1}
    The optimum transmission policy for $N>2$  is a threshold policy represented by the quantities $\tau_{ij}$, i.e., the source transmits when the AoII process exceeds $\tau_{ij}$, when $X(t)=i$ and $\hat{X}(t)=j$, 
\end{lemma}
\begin{Proof}
First, it is clear that when $X(t)=\hat{X}(t)$, transmitting a packet has no effect except the additional sampling cost. Therefore, we are only interested in cases when $X(t)\neq\hat{X}$(t). For state $s=(i,j,\Delta,0)$ and given any $\lambda$, the decision rule $g^\lambda_{ij}(\Delta)$ that indicates whether the transmission is optimum or not is derived in Appendix B. For any $\Delta'>\Delta$, the difference between decision rules $g^\lambda_{ij}(\Delta')-g^\lambda_{ij}(\Delta)$ is bounded as
\begin{align}
&g^\lambda_{ij}(\Delta')-g^\lambda_{ij}(\Delta)\geq 0
\end{align}
Notice that this bound holds for any $\lambda$ including to $\lambda^*$ in Appendix~A that finds the optimum policy for the unconstrained problem. Therefore, if starting transmission is optimum when $X(t)=i$, $\hat{X}=j$, and AoII$=\Delta$, it is also optimum for any age value greater than $\Delta$. In other words, there exist optimum thresholds $\tau_{ij}$ for starting transmission for $i \neq j$, and the preemption is never optimum unless $X(t)$ changes.
\end{Proof}
\begin{remark} \label{rem:N2}
    Symmetric Markov processes with $N$ states can be expressed with a generator matrix $Q=\{q_{ij}\}$ where $q_{ij}=\frac{\sigma}{N-1}$ for $i \neq j$ and $q_{ii}=-\sigma$. It is worth noting that because of this symmetry among states, all optimum threshold values are equal to each other as $\tau_{ij}=\tau$ for $i \neq j$, and $\tau_{ii}=0$ for symmetric processes. Moreover, for the case of binary sources (not necessarily symmetric), there are only two non-zero thresholds corresponding to $(X(t)=1,\hat{X}(t)=2)$ and $(X(t)=2,\hat{X}(t)=1)$. 
\end{remark}

\section{Relaxed System}  \label{sec:4}
In this section, we focus on a relaxed scenario for which the source waits for the AoII process to reach a threshold  $\tau_j$ when $\hat{X}(t)=j$ for transmitting its status update packets. Note from Lemma~\ref{lem:1} and Remark~\ref{rem:N2} that the policy obtained for the relaxed problem is optimum for $N=2$ and also for the case CTMC is symmetric. For the purpose of constructing the CSMDP model, we first define a synchronization state (SS) $S_j, j\in \mathcal{N}$ of this SMDP at a synchronization point (SP) $t$ if $X(t)$ and $\hat{X}(t)$ are just synchronized at the value $j$, i.e., $X(t)=\hat{X}(t)=j$, $X(t^-) \neq \hat{X}(t^-)$. Each SS is only related to the value of $\hat{X}(t)$, thus the state space is reduced to $\mathcal{S}=\mathcal{N}$. The action taken at state $S_j$ at the corresponding SP is the threshold value $\tau_j\in{\rm I\!R}_+$. Finally, we can adapt the age cost $A_j(\tau_j)$, sampling cost $C_j(\tau_j)$  and sojourn time $D_j(\tau_j)$, which are all defined in Section~\ref{sec:2} with their expected values $a_j(\tau_j)$, $c_j(\tau_j)$ and $d_j(\tau_j)$, respectively. The CSMDP model is then completed with the calculation of the transition probabilities $p_{ji}(\tau_j)$ from $S_j$ to $S_i$ when threshold $\tau_j$ is enforced. 

The optimum threshold values are found by solving the average cost CSMDP problem using policy iteration \cite{ibe2013markov,tomasevicz2006optimum} in conjunction with the Lagrangian approach \cite{makowski1986}. Similar to the general setting, we consider a constrained primal problem, and an unconstrained Lagrangian problem, which are adapted versions of Eqn.~\eqref{Optc} and Eqn.~\eqref{OptUc} in Appendix~A to the relaxed problem, respectively. The policy iteration algorithm in \cite{ibe2013markov,tomasevicz2006optimum} is used to solve the unconstrained problem in Algorithm~\ref{alg:cap} for any given Lagrangian multiplier $\lambda$. In order to find the optimum thresholds to minimize the constrained problem, first the Algorithm~\ref{alg:cap} is solved for $\lambda=0$. If the sampling rate (which can be calculated by Eqn.~\eqref{eq:rate_app} in Appendix~D for the obtained thresholds), is less than the budget $b$, then we have the optimum solution. Otherwise, the optimum Lagrange multiplier $\lambda^*$ with the corresponding thresholds satisfying the constraint on the boundary, is found by the bisection search algorithm with the sensitivity parameter $\epsilon_\lambda$. This procedure is summarized in Algorithm~\ref{alg:lag} in Appendix~D.

\begin{algorithm}
\caption{Policy iteration algorithm}\label{alg:cap}
\begin{algorithmic}
\State \textbf{Initialize:} $\tau_{j}=0$ for $j\in \mathcal{N}$.
\State \textbf{Step 1: (CSMDP model)}  Obtain the values $a_j(\tau_j)$, $c_j(\tau_j)$, $d_j(\tau_j)$, and $p_{ji}(\tau_j)$ for given vector of thresholds $\tau_j$. 
 \State \textbf{Step 2: (value determination)}: Obtain the long-time average cost $\eta$ and the relative values $v_j$ for $1 \leq j <N$ by fixing $v_N=0$ and solving the following $N$ optimality equations 
    \begin{align}
 v_j & =a_j(\tau_j)+\lambda c_j(\tau_j) -\eta d_j(\tau_j)+\sum_{i=1}^N p_{ji}(\tau_j)v_j.
    \end{align}
    \State \textbf{Step 3: (policy improvement)}: For each $j$, set $\tau_j$ to 
    \begin{align}
          &  \arg \min   \dfrac{a_{j}(\tau_j)+\lambda c_j(\tau_j)}{d_j(\tau_j)}+\dfrac{\sum_{j\neq i} (v_{j}-v_i) p_{ij}(\tau_j) }{d_j(\tau_j)}, \label{eq:pol}
    \end{align}
    \State \textbf{Step 4: (stopping rule)} If $|\eta^{(n)}-\eta^{(n-1)}|\leq \epsilon_\eta$ then stop. Otherwise go to Step 1.  Here, $\eta^{(n)}$ denotes the long-time average cost obtained at iteration $n$.
\end{algorithmic}
\end{algorithm}

\begin{figure}[t]
    \centering
   \includegraphics[width=0.85\columnwidth]{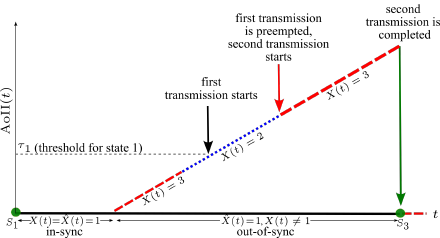} 
    \caption{A sample path for $X(t)$, $\hat{X}(t)$ and $\text{AoII}(t)$ for an example scenario. Green circles denote the synchronization points.}
    \label{fig:trans2}
\end{figure}

In the remainder of this section, we will brief our main results related to Step~1 of Algorithm~\ref{alg:cap} with the detailed proofs provided in the Appendix section. For this purpose, a sample path is illustrated in Fig.~\ref{fig:trans2} for $N=3$ starting from $S_1$ and $\tau_1$ is the threshold value, thus the source stays idle until $\text{AoII}(t)$ reaches the value $\tau_1$ upon which the source starts transmitting the current observation, i.e., state 2, which requires an exponentially distributed service time with parameter $\mu$. However, before the transmission is complete, a state transition from state 2 to state 3 takes place. Therefore, the source preempts the ongoing transmission and re-samples $X(t)$ while at state 3. Finally, the system reaches the next synchronization state $S_3$ at which point $X(t)$ and $\hat{X}(t)$ are synchronized back at value 3, since this second transmission has completed without facing a state change in the source process. Notice that it is possible that we can reach the next SS $S_1$ by a state change of the original process $X(t)$ from states 2 or 3 to the original state 1, before or after it reaches the threshold value, but this situation is not illustrated in Fig.~\ref{fig:trans2}.

The material below is based on \cite{kemeny1960finite}. An absorbing MC (AMC) $Y(t)$ with $K$ transient and $L$ absorbing states has a generator of the form 
\begin{align}
 \left[ \begin{array}{c:c}
   \bm A & \bm B \\
   \hdashline
   \bm{0} & \bm{0} \\
\end{array}\right], \label{eq:AMC}
\end{align}
where $\bm A_{K \times K}$ and $\bm B_{L \times L}$ correspond to the transition rates among the transient states, and from the transient states to the absorbing states, respectively, and $\bm{\beta} = \{ \beta_i \}$ is a ${1 \times K}$ row vector with $\beta_i$ denoting the initial probability of being in transient state $i$. In this case, we say $Y(t)$ is an AMC characterized with the triple $(\bm A,\bm B,\bm{\beta})$, i.e., $Y(t) \sim AMC(\bm A,\bm B,\bm{\beta})$. The time required until absorption for the AMC $Y(t)$, denoted by $\Theta$, is said to have a PH-type distribution characterized with the pair $(\bm A,\bm{\beta})$, i.e., $\Theta \sim PH(\bm A,\bm{\beta})$ with its probability density function (pdf) being in the form $f_{\Theta}(x) = -\bm{\beta}\mathrm{e}^{\bm{A}x}\bm{A}\bm{1}$, where $\bm{1}$ is a column vector of ones of appropriate size \cite{latouche1999introduction}.

For the CSMDP model, the process starting from SS $S_j$ until the next SS can be modeled by means of two AMCs $Y_{1,j}(t)$ and $Y_{2,j}(t)$, with their time to absorption, denoted by $T_{1,j}$ and $T_{2,j}$, respectively. For the AMC $Y_{1,j}(t)$, we have $N-1$ transient states corresponding to $X(t)=i\neq j$, and one absorbing state, which is $S_j$. While the transition rates from state $i$ to $i'$ for $i \neq i'$ is $q_{ii'}$, the absorption rate from $i$ to $S_j$ is $q_{ij}$. The process starts from state $i$ with probability ${q_{ji}}/{\sigma_j}$. Therefore, $Y_{1,j}(t) \sim AMC(\bm{A_1}^{(j)},\bm{B_1}^{(j)},\bm{\beta_1}^{(j)})$ where
\begin{align}
    \bm{A_1}^{(j)} & =\bm{Q}^{(-j)}, \quad \bm{B_1}^{(j)}=\bm{q}^{(j)}, \quad    \bm{\beta_1}^{(j)} = \dfrac{1}{\sigma_j}\bm{q_r}^{(j)}, 
\end{align}
where $\bm{Q}^{(-j)}$ is $\bm{Q}$ but its $j$th row and $j$th column removed, and $\bm{q}^{(j)}$ is the $j$th column of $\bm{Q}$ excluding $q_{jj}$, and  $\bm{q}_r^{(j)}$ is the $j$th row of $\bm{Q}$ excluding $q_{jj}$. If $Y_{1,j}(t)$ is not absorbed within a duration of $\tau_j$, the next AMC process $Y_{2,j}(t)$ starts which has $N-1$ transient states corresponding to $X(t)=i\neq j$, and $N$ absorbing states corresponding to each SS. While all transitions among transient states, and absorption rates to $S_j$ are the same as the AMC $Y_{1,j}(t)$, the absorption rates to SS $S_i$ equals to $\mu$ for $i\neq j$. Therefore, 
$Y_{2,j}(t) \sim AMC(\bm{A_2}^{(j)},\bm{B_2}^{(j)},\bm{\beta_2}^{(j)})$ where 
\begin{align}
    \bm{A_2}^{(j)} & =\bm{Q}^{(-j)}-\mu \bm{I}, \quad \bm{B_2}^{(j)}=\begin{bmatrix} \bm{q}^{(j)} & \mu\bm{I} \end{bmatrix},
\end{align}
where $\bm{I}$ is an identity matrix of appropriate size. Additionally, the initial probability vector for $Y_{2,j}(t)$ equals to the normalized state probabilities in $Y_{1,j}(t)$ after a duration of $\tau_j$. Consequently, we write
\begin{align}
        \bm{\beta_2}^{(j)} &=\frac{1}{1-\kappa_j} \bm{\beta}_1^{(j)}\mathrm{e}^{\tau_j\bm{A}_1^{(j)}},
\end{align}
where $\kappa_j$ is the absorption probability within a duration of $Y_{1,t}$ and hence it can be written as,
\begin{align}
    \kappa_j=\mathbb{P}[T_{1,j}<\tau_j]=1-\bm{\beta}_1^{(j)}\mathrm{e}^{\tau_j\bm{A}_1^{(j)}}\bm{1}.
\end{align}
Returning to the CSMDP formulation, we are now ready to express $D_j(\tau_j)$ and $A_j(\tau_j)$ as, 
\begin{align}
    D_j(\tau_j) & =\begin{cases}
        T_{1,j}+H_j & T_{1,j}<\tau_j, \\
        \tau_j+T_{2,j}+H_j &  \text{otherwise}, 
    \end{cases} \\
    A_j(\tau) & =\dfrac{(D_j(\tau)-H_j)^2}{2},
\end{align}
whose expected values can be written using the unconditional and certain conditional moments of phase-type distributions in Appendix~C, as 
\begin{align}
 d_j(\tau)  &=\bm{\beta}_1^{(j)} \Bigl( (\bm{U_1}^{(j)}-\bm{U_2}^{(j)})\mathrm{e}^{\tau_j \bm{A}^{(j)}_1} \Bigr) \bm{1}- \bm{\beta}_1^{(j)} \bm{U_1}^{(j)} \bm{1}  + 1/\sigma_j \\
    a_j(\tau)&=\tau_j \bm{\beta}_1^{(j)}\mathrm{e}^{\tau_j \bm{A}^{(j)}_1}(\bm{U_1}^{(j)}-\bm{U_2}^{(j)})\bm{1}+\bm{\beta}_1^{(j)}\bm{V_1}^{(j)}\bm{1} \nonumber\\
    &\quad+\bm{\beta}_1^{(j)}\mathrm{e}^{\tau_j \bm{A}^{(j)}_1}\left(\bm{V}^{(j)}_2-\bm{V}^{(j)}_1\right)\bm{1},  
\end{align}
where $\bm{U_k}^{(j)}=(\bm{A_k}^{(j)})^{-1}$ and $\bm{V_k}^{(j)}=(\bm{A_k}^{(j)})^{-2}$ for $k=1,2$.

In order to obtain $c_j(\tau_j)$, we need to find the average number of times a transient state is visited before absorption of the AMC $Y_{2,j}(t)$. By using the characteristic matrix definition in Appendix~C, it can be expressed as
\begin{align}
    c_j(\tau_j)=\bm{\beta}_1^{(j)} \mathrm{e}^{\tau_j\bm{A}_1^{(j)}} (\bm I-\bm D^{(j)})^{-1} \bm{1}, \label{eq:rj}
\end{align}
where $\bm{D}^{(j)}$ is the probability transition matrix of a DTMC obtained at the embedded epochs of state transitions of the generator $\bm{A_2}^{(j)}$. Note that the transition from $S_j$ to $S_i$ for $j\neq i$ occurs if the process $Y_{2,j}(t)$ is reached, and it is absorbed by $S_j$. With the help of equations in Appendix~C, this probability can be calculated as
\begin{align}
    p_{ji}(\tau_j)&=-\bm{\beta}_1^{(j)} \mathrm{e}^{\tau_j\bm{A}_1^{(j)}} \bm{A}_2^{(j)} \bm{B}_2^{(j)}\bm{e}^{(j)}, \label{eq:pji}
\end{align}
where $\bm{e}^{(j)}$ is a column vector of zeros of appropriate size except for its $j$th element which is one.
On the other hand, the self-transition probability for $S_j$ is written as,
\begin{align}
    p_{jj}(\tau_j)=-\bm{\beta}_1^{(j)} \mathrm{e}^{\tau_j\bm{A}_1^{(j)}} \bm{A}_2^{(j)} \bm{B}_2^{(j)}\bm{e}^{(j)}+\kappa_j. \label{eq:pjj}
\end{align}
For the complexity of the proposed algorithm, the mathematical operation with the highest complexity in the proposed algorithm is matrix inversion of the matrices $\bm{A}_1^{(j)}$ and $\bm{A}_2^{(j)}$, which has a complexity of at most $\mathcal{O}(N^3)$. Moreover, these operations are repeated $N$ times in each iteration of the policy iteration algorithm making the overall complexity $\mathcal{O}(N^4)$, which is also confirmed numerically as $N$.

\section{Numerical Results}  \label{sec:5}
\begin{figure}[t]
    \centering
    \includegraphics[width=0.85\columnwidth]{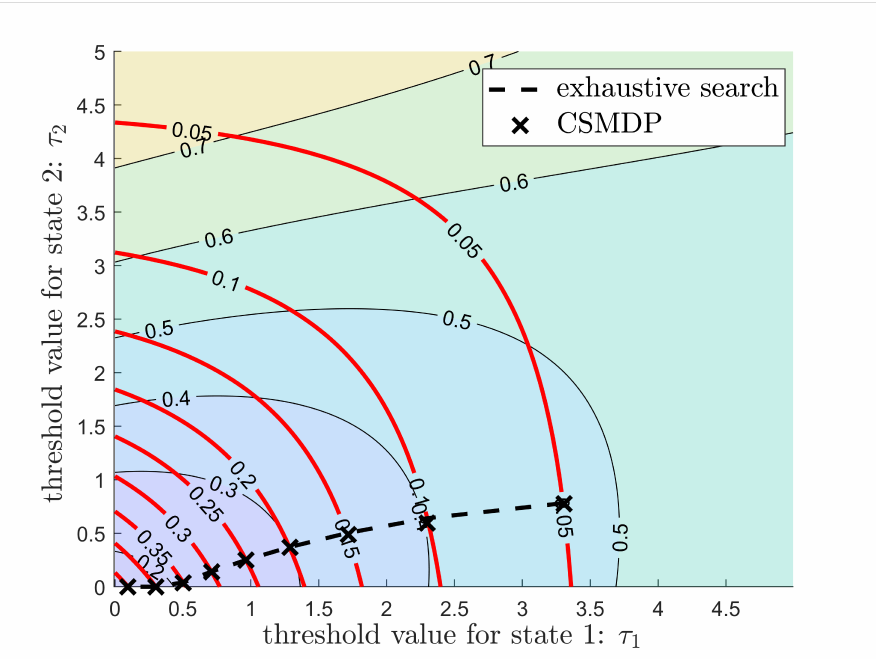}
    \caption{The red (resp. gray) contour lines correspond to the threshold pairs with the same sampling rate $R$ (resp. same MAoII). The minimum MAoII points for a given sampling rate $R$ are connected with a dashed line, whereas the optimum threshold pairs obtained by the proposed CSMDP are marked by a cross.}
    \label{fig:exh2d}
\end{figure}

In the numerical examples, the sensitivity parameters for the proposed method are chosen as $\epsilon_\lambda=\epsilon_\eta=10^{-2}$.
Two different CTMCs are considered in the numerical examples with their generators given below,
\begin{align}
 \bm{Q}_1= \begin{bmatrix}
        -.6 & .6 \\ .75 & -.75
    \end{bmatrix},     \bm{Q}_2&= \begin{bmatrix}
        -1.025 & 1 & .025 \\
        .05 & -.75 & .7 \\
        .4 & .01 & -.41
    \end{bmatrix} \label{sources23}
\end{align}
In Fig.~\ref{fig:exh2d}, we set $\mu=1$ and contour lines are presented for the binary Markov source with generator $\bm{Q}_1$ for the threshold pair $(\tau_1,\tau_2)$ using mathematical analysis. The red (resp. gray) contour lines correspond to the threshold pairs with the same sampling rate $R$ (resp. same MAoII). The minimum MAoII points for a given sampling rate $R$ are connected with a dashed line, whereas the optimum threshold pairs obtained by the proposed CSMDP are marked by a cross, which shows perfect agreement between the two, validating the proposed algorithm.  

\begin{figure}[t]
    \centering
    \includegraphics[width=0.9\columnwidth]{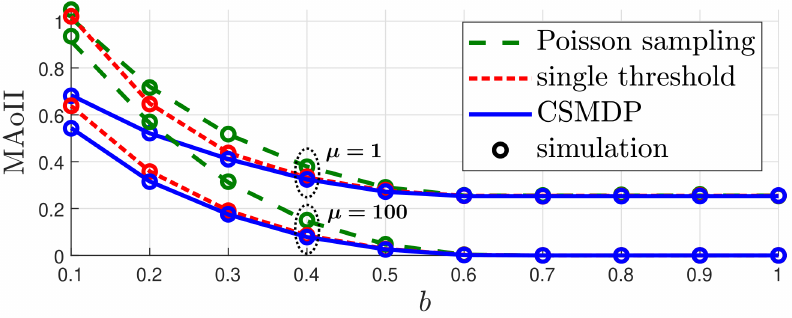}
    \caption{MAoII depicted as a function of the sampling rate budget $b$ for the generator $\bm{Q}_2$, using three policies. 
    }
    \label{fig:sim1}
\end{figure}

In Fig.~\ref{fig:sim1}, MAoII performance of the CSMDP approach is compared with two baseline policies for $\bm{Q}_2$, when the sampling budget $b$ is varied for two values of $\mu$. The first baseline policy is the \emph{single threshold} policy when all thresholds are identical and the optimum threshold for this policy is obtained with one-dimensional exhaustive search. The second baseline policy is \emph{Poisson sampling} for which the sensor is to sample the source process with Poisson intensity $\gamma$ when $X(t) \neq \hat{X}(t)$. The value of $\gamma$ under which the budget is met is found by one-dimensional search using the analytical model of \cite{cosandal2024modeling}. For all cases, we verify our analytical results by running the simulations for $10^5$ transmissions. As the budget $b\rightarrow \infty$, the MAoII performances of the three schemes turn out to be the same but the error floor for the $\mu=1$ case stems from moderate packet service times. For lower values of the budget parameter $b$, the CSMDP approach significantly outperforms the two baseline policies. 

\section{Conclusions}  \label{sec:6}
We investigated AoII-optimum transmission policies under sampling rate constraints for the tracking of a CTMC source process at a remote monitor. We first showed that the optimum policy is of threshold type where the thresholds depend on the instantaneous values of both the source and monitor processes. Then, we considered a relaxed problem where the thresholds are allowed to be dependent only on the monitor process and we proposed a CSMDP solution for this problem using policy iterations and the Lagrangian approach, with a computational complexity of $\mathcal{O}(N^4)$. Numerical examples are presented to validate the proposed method which is shown to substantially outperform the two baseline policies especially when the sampling rate budget is relatively low. 

\clearpage 
\bibliographystyle{unsrt}
\bibliography{bibl}
%\clearpage
\appendices
\section{Lagrangian Method} \label{app:dual}
The Lagrangian method is used to solve a constrained MDP by converting an unconstrained MDP with a Lagrangian multiplier $\lambda$ and iteratively solving it for different values of the Lagrangian multiplier $\lambda$ \cite{makowski1986}. This method can be adapted for a CSMDP as follows. First consider the constrained problem 
\begin{mini}
	{\phi}{J^{\phi} } 
	{\label{Optc}}
    {}
	\addConstraint{ K^{\phi} }{\leq b} 
\end{mini}
where $J^\phi$ and $K^\phi$ are, respectively, the average cost objective function (average AoII in our case) and the constraint function (the long term sampling rate in our case) for a feasible policy $\phi$. If $s_n \in \mathcal{S}$ and $u_n \in \mathcal{U}$ are the state and action, respectively, at the $n$th decision epoch for policy $\phi$, then the functions $J^\phi$ and $K^\phi$ can be expressed as
\begin{align}
   J^\phi&=\lim_{L\to\infty} \dfrac{ \mathbb{E}^\phi \left[\sum_{n=1}^L A_{s_n}(u_n) \right] }{ \mathbb{E}^\phi \left[\sum_{n=1}^L D_{s_n}(u_n)\right] } \\
    K^\phi&= \lim_{L\to\infty} \dfrac{ \mathbb{E}^\phi \left[\sum_{n=1}^L C_{s_n}(u_n) \right] }{ \mathbb{E}^\phi \left[\sum_{n=1}^L D_{s_n}(u_n)\right] },
\end{align}
where $A_{s}(u)$ (resp. $C_{s}(u)$) is the total cost (resp. constraint cost) during the sojourn time $D_{s}(u)$ of the 
time interval starting from state $s$ while applying the action $u$.

The constrained problem can be transformed into a Lagrangian unconstrained problem as
\begin{mini}
	{\phi}{J^{\phi}+\lambda K^\phi} 
	{\label{OptUc}}
    {}
\end{mini}
where $J^\phi+\lambda K^\phi$ can be expressed as
\begin{align}
    J^\phi+\lambda K^\phi=\lim_{L\to\infty} \dfrac{ \mathbb{E}^\phi \left[\sum_{n=1}^L A_{s_n}(u)+\lambda C_{s_n}(u) \right] }{ \mathbb{E}^\phi \left[\sum_{n=1}^L D_{s_n}(u_n)\right] },
\end{align}
In the same work, it is shown that there exists a Lagrangian coefficient $\lambda^*$ such that, optimum policy $\phi^*$ obtained from the unconstrained problem is also optimum if it satisfies $K^{\phi^*}=b$, or $\lambda=0$ and $K^{\phi^*}\leq b$.

\section{On the optimality of the threshold policy} \label{app:proof_opt}

The optimality equations \cite{ibe2013markov,white1993mdp} for the unconstrained Lagrangian problem defined in Eqn.~\eqref{Optc} can be expressed  as
\begin{align}
    v_s&=\min_{u\in\mathcal{U}} v_s(u)\\
    &=\min_{u\in\mathcal{U}}\{a_s(u)+\lambda c_s{u}-\eta d_{s}(a)+\sum_{s' \in \mathcal{S}}^N p_{s,s'}(u) v_{s'}(a)\}, \nonumber\label{eq:opt_gen}
\end{align}
where $v_s$ and $v_s(u)$ are the value of state $s$ for optimum policy and feasible policy $u$, respectively, $\eta$ corresponds to the long-time average reward. Additionally, $a_s(u)$, $d_s(u)$ and $c_s(u)$ are the expected values of $A_s(u)$, $D_s(u)$ and $C_s(u)$, respectively.

Now consider that state $s=(i,j,\Delta,0)$ for $i \neq j$. For $u=0$ is chosen, there are three possibilities which are summarized in Table~\ref{tab:u0}. First, the process $X(t)$ does not change and the age increases by $\delta$ with the probability of $\mathbb{P}(H_i>\delta)$, second the process changes to state $X(t)=j$ and the AoII becomes 0 with a probability of $p_{ij}\mathbb{P}[H_i>\delta]$, and third the process $X(t)$ changes to state $X(t)=k$ which $k\neq j$ and the age increases by $H_i$ with the probability of $p_{ik}\mathbb{P}[H_i>\delta]$.  
By considering these probabilities and age values, we can write $v_{i,j,\Delta,0}(0)$ as

\begin{align} 
    &= \mathbb{P}[H_i>\delta] \left( \mathbb{E}\left[\frac{\delta^2}{2}+\Delta \delta\bigg|H_i>\delta\right]+ v_{ i,j,\Delta+\delta ,0} -\eta \delta \right) \nonumber  \\
    &+p_{ij}\mathbb{P}[\delta>H_i]\left( \mathbb{E}\left[\dfrac{H_i^2}{2}+\Delta H_i \bigg|\delta>H_i\right]+ v_{ j,j,0 ,0} -\eta H_i \right) \nonumber \\
    &+\sum_{k\neq j} p_{ik}\mathbb{P}[\delta>H_i] \bigg( \mathbb{E}\left[\dfrac{H_i^2}{2}+\Delta H_i \bigg|\delta>H_i\right]\nonumber \\
    &+ E[v_{ k,j,\Delta+H_i ,0}|H_i<\delta]-\eta H_i \bigg). 
\end{align}
By using the law of total expectation, we can write it as
\begin{align}
    v_{i,j,\Delta,0}(0)&=\mathbb{P}[H_i>\delta]  v_{ i,j,\Delta+\delta ,0}+p_{ij}\mathbb{P}[\delta>H_i] v_{ j,j,0 ,0}) \nonumber
    \\&+\sum_{k\neq j} p_{ik}\mathbb{P}(\delta>H_i)  E[v_{ k,j,\Delta+H_i ,0}|H_i<\delta]\nonumber\\
    &+E[M_0(\Delta-\eta+M_0)]
    \label{eq:a0short}
\end{align}
where $M_0=\min(H_i,\delta)$.

Similarly, we can repeat the same for $u=1$, and all relevant parameters are summarized in Table~\ref{tab:u1}, and the corresponding value can be expressed as
\begin{align}
    v_{i,j,\Delta,0}(1)&=\mathbb{P}[H_i,E>\delta]   v_{ i,j,\Delta+\delta,1} +p_{ij}\mathbb{P}[E,\delta>H_i]v_{ j,j,0 ,0}  \nonumber \\
    &+\sum_{k\neq j}p_{ik}\mathbb{P}[E,\delta>H_i] v_{ k,j,\Delta+H_i ,0} \nonumber \\
    &+\mathbb{P}[\delta,H_i>E] v_{ i,i,0 ,0}+\mathbb{E}[M_1(\Delta-\eta+M_1)] + \lambda,
    \label{eq:a1short}
\end{align}
where $M_1=\min(H_i,E,\delta)$.

Now, by combining equations \eqref{eq:a0short} and \eqref{eq:a1short}, we can write the decision rule for optimum action $u^*$ as
\begin{align}
g_{ij}(\Delta)&=\mathbb{E}[M_0(M_0+\Delta-\eta)]-\mathbb{E}[M_1(M_1+\Delta-\eta)] \nonumber\\
&+\mathbb{P}[H_i>\delta] v_{ i,j,\Delta+\delta ,0} - \mathbb{P}[H_i,E>\delta] v_{ i,j,\Delta+\delta ,1})\nonumber \\
&+(\mathbb{P}[\delta>H_i]-\mathbb{P}(\delta,E>H_i))p_{ij} v_{ j,j,0 ,0} \nonumber\\
&+ \sum_{k\neq i,j} p_{ik} \mathbb{E}[    v_{ i,j,\Delta+H_i ,0}|H_i<\delta]\mathbb{P}[H_i<\delta]  \nonumber\\
&-\sum_{k\neq i,j} p_{ik} \mathbb{E}[ v_{ i,j,\Delta+H_i ,0}|H_i<\delta,E] \mathbb{P}[H_i<\delta,E] \nonumber \\
&- \mathbb{P}[H_i,\delta>E] v_{i,i,0,0} -\lambda  \underset{u^*=0}{\overset{u^*=1}{\gtrless}} 0. \label{eq:dec}
\end{align}

Now notice that
\begin{itemize}
    \item $v_{i,j,\Delta+d,0} \geq  v_{i,j,\Delta+d,1}$, because if there is an ongoing transmission the sensor can preempt it.
    \item $\mathbb{P}[\delta>H_i]-\mathbb{P}[\delta,E>H_i]$ and $\mathbb{P}[H_i>\delta]-\mathbb{P}[E,H_i>\delta]$ are always positive, and independent from $\Delta$.
    \item $\mathbb{E}[M_0)-\mathbb{E}[M_1]$ is always positive.
    \item $v_{i,j,\Delta}$ is always non-negative and monotonically non-decreasing with $\Delta$ because Lagrangian cost is always monotonically non-decreasing with $\Delta$ as it can be seen from Tables~\ref{tab:u0}~and~\ref{tab:u1}.
\end{itemize}

Then for $\Delta'>\Delta$, we can express $g_{ij}(\Delta')-g_{ij}(\Delta)$ as
\begin{align}
    &g_{ij}(\Delta')-g_{ij}(\Delta)=(\Delta'-\Delta)(\mathbb{E}[M_0]-\mathbb{E}[M_1]) \nonumber\\
    & +\mathbb{P}[H_i>d] (v_{ i,j,\Delta'+\delta ,0}-v_{ i,j,\Delta+\delta ,0}) \nonumber\\
    &-\mathbb{P}[H_i,E>d] (v_{ i,j,\Delta'+\delta ,1}-v_{ i,j,\Delta+\delta ,1} \nonumber\\
    &+ \sum_{k\neq i,j} p_{ik}  E[v_{ i,j,\Delta'+H_i ,0}-v_{ i,j,\Delta+H_i ,0}|H_i<\delta]\mathbb{P}[H_i<\delta]   \nonumber\\
    &+ \sum_{k\neq i,j} p_{ik}  E[v_{ i,j,\Delta'+H_i ,0}-v_{ i,j,\Delta+H_i ,0}|H_i<\delta,E]\mathbb{P}[H_i<\delta,E]   \nonumber\\
    &\geq (\Delta'-\Delta)(E[M_0]-E[M_1]) \nonumber\\
    & +\big(\mathbb{P}[H_i>\delta]-\mathbb{P}[H_i,E>\delta]\big) (v_{ i,j,\Delta'+\delta ,0}-v_{ i,j,\Delta+\delta ,0})\nonumber\\
    &+ \big(P[\delta>H_i]-\mathbb{P}[\delta,E>H_i]\big) \nonumber\\
    &\sum_{k\neq i,j} p_{ik}  E[v_{ i,j,\Delta'+H_i ,0}-v_{ i,j,\Delta+H_i ,0}|H_i<\delta] \geq 0
\end{align}

Notice that this bound holds for any $\lambda$ including to $\lambda^*$ that finds the optimum policy for the main problem.

\begin{table*}[t]
\centering
\caption{ SMDP parameters for state $s=(i,j,\Delta,0)$, and action $u=0$ } \label{tab:u0}
\begin{tabular}{|c|c|c|c|} 
\hline
                 next state            &   condition                           &   Lagrangian cost                             &   sojourn times\\ \hline
{$(i,j,\Delta+\delta,0)$} & {$\delta<H_i$}            & {$\frac{\delta^2}{2}+\Delta \delta$} & {$\delta$} \\ \hline
{$(i,j,\Delta+H_i,0)$} & {$H_i<\delta, \ i \to k$} & {$\frac{H_i^2}{2}+\Delta H_i$} & {$H_i$} \\ \hline
{$(j,j,0,0)$}        & {$H_i<\delta, \ i \to j$} & {$\frac{H_i^2}{2}+\Delta H_i$} & {$H_i$}  \\ \hline
\end{tabular}
\end{table*}

\begin{table*}[t]
\centering
\caption{ SMDP parameters for state $s=(i,j,\Delta,0)$, and action $u=1$ } \label{tab:u1}
\begin{tabular}{|c|c|c|c|} 
\hline
                 next state            &   condition                           &   Lagrangian cost                             &   sojourn times  \\ \hline
{$(i,j,\Delta+\delta,1)$} & {$\delta<H_i,E$}            & {$\frac{\delta^2}{2}+\Delta \delta + \lambda$ } & {$\delta$} \\ \hline
{$(i,j,\Delta+H_i,0)$} &  {$H_i<\delta,E, \ i \to k$} & {$\frac{H_i^2}{2}+\Delta H_i + \lambda$} & {$H_i$} \\ \hline
{$(j,j,0,0)$}        & {$H_i<\delta,E, \ i \to j$} & {$\frac{H_i^2}{2}+\Delta H_i+\lambda$} & {$H_i$}  \\ \hline
{$(i,i,0,0)$}        & {$E<H_i,\delta$} & {$\frac{E^2}{2}+\Delta E+\lambda$} & {E}  \\ \hline
\end{tabular}
\end{table*}
\section{Absorbing Markov chains and Phase-type distribution} \label{app:abs}
The following is needed to follow the proposed analytical method for absorbing Markov chains (AMC) which refers to continuous-time Markov chains with absorbing states. Consider a AMC $Y(t)$ with $K$ transient and $L$ absorbing states. Let the generator of this CTMC be written as,
\begin{align}
 \left[ \begin{array}{c|c}
   \bm A & \bm B \\
   \midrule
   \bm{0} & \bm{0} \\
\end{array}\right] \label{eq:Tmatrix}
\end{align}
where $\bm A_{K \times K}$ and $\bm B_{K \times L}$ correspond to the transition rates among the transient states, and from the transient states to the absorbing states, respectively. In this case, we say $Y(t)$ is an AMC characterized with the triple $(\bm A,\bm B,\bm{\beta})$, i.e., $Y(t) \sim AMC(\bm A,\bm B,\bm{\beta})$, where $\bm{\beta} = \{ \beta_i \}$ is a ${1 \times K}$ row vector with $\beta_i$ denoting the initial probability of being in transient state $i$. 

Upon merging all the absorbing states into one, the distribution of time until absorption, denoted by $T$, is known as the phase-type distribution \cite{latouche1999introduction} whose probability density function (PDF) and cumulative density function (CDF) are written in terms of the matrix exponential function of $t\bm A$ as,
\begin{align}
    f_T(t)=-\bm{\beta} \mathrm{e}^{t\bm A}\bm{A}\bm{1}, \quad F_T(t)=1-\bm{\beta} \mathrm{e}^{t\bm A}\bm{1},
    \label{eq:pdf}
\end{align}
from which the first and second moments can be obtained,
\begin{align}
    \mathbb{E}[T]=-\bm{\beta} \bm A^{-1} \bm{1}, \quad
    \mathbb{E}[T^2]=2\bm{\beta} \bm A^{-2} \bm{1}, \label{eq:mom12}
\end{align}
where $\bm{1}$ denotes a column vector of ones of appropriate size. Similarly, we can write the conditional first moment $\mathbb{E}[T  |  T<\tau]$  as (also see \cite{cai2005conditional}) 
\begin{align}
    &\mathbb{E}[T|T<\tau]=\dfrac{-\tau\bm{\beta}e^{\tau\bm{A}}\bm{A}^{-1}\bm{1}+\bm{\beta}e^{\tau\bm{A}}\bm{A}^{-1}\bm{1}-\bm{\beta}\bm{A}^{-1}\bm{1} }{1-\bm{\beta}e^{\tau\bm{A}}\bm{1}},\label{eq:momc1}  
\end{align}
and the conditional second moment $E[T_{1,j}^2 |  T_{1,j}<\tau_j]$ as
\begin{align}
     &= \dfrac{-\tau^2\bm{\beta}e^{\tau\bm{A}}\bm{1}+2\tau\bm{\beta}e^{\tau\bm{A}}\bm{A}^{-1}\bm{1}-2\bm{\beta}\bm{A}^{-2}\bm{1}+2\bm{\beta}\bm{A}^{-2}\bm{1} }{1-\bm{\beta}e^{\tau\bm{A}}\bm{1}}
     \label{eq:momc2}  
\end{align}

By using these identities, we can calculate variables $a_j(\tau_j)$, $d_j(\tau_j)$ in the manuscript as following,
\begin{align}
      d_j(\tau)&=     F_{T_{1,j}}(\tau_j)\mathbb{E}[T_{1,j}|T_{1,j}<\tau] \nonumber \\
     &+\left(1-F_{T_{1,j}}(\tau_j)\right)(\tau_j+\mathbb{E}[T_{2,j}])+\frac{1}{\sigma_j} \label{eq:dj}  \\
    &=\bm{\beta}_1^{(j)} \Bigl( (\bm{U_1}^{(j)}-\bm{U_2}^{(j)})\mathrm{e}^{\tau_j \bm{A}^{(j)}_1} \Bigr) \bm{1}- \bm{\beta}_1^{(j)} \bm{U_1}^{(j)} \bm{1}  + 1/\sigma_j \\ 
    a_j(\tau)&=F_{T_{1,j}}(\tau_j)(\tau_j)\mathbb{E}[T_{1,j}^2|T_{1,j}<\tau_j] \nonumber\\
    &+\left(1-F_{T_{1,j}}(\tau_j)\right)\left(\dfrac{\tau_j^2}{2}+\tau \mathbb{E}[T_{2,j}] +\dfrac{\mathbb{E}[T_{2,j}^2]}{2}\right) \nonumber \\
    &=\tau_j \bm{\beta}_1^{(j)}\mathrm{e}^{\tau_j \bm{A}^{(j)}_1}(\bm{U_1}^{(j)}-\bm{U_2}^{(j)})\bm{1}+\bm{\beta}_1^{(j)}\bm{V_1}^{(j)}\bm{1} \nonumber\\
    &+\bm{\beta}_1^{(j)}\mathrm{e}^{\tau_j \bm{A}^{(j)}_1}\left(\bm{V}^{(j)}_2-\bm{V}^{(j)}_1\right)\bm{1}, \label{eq:aj}
\end{align}
where $\bm{U_k}^{(j)}=(\bm{A_k}^{(j)})^{-1}$ and $\bm{V_k}^{(j)}=(\bm{A_k}^{(j)})^{-2}$ for $k=1,2$, and other parameters are specified in the original manuscript.

 Moreover, the probability of being absorbed in absorbing state $j$ is,
\begin{align}
    p_{j} & =-\bm{\beta} \bm{A}^{-1} \bm{B} \bm{e}^{(j)}, \label{eq:prob}
\end{align}
where $\bm{e}^{(j)}$ is a column vector of zeros of appropriate size except for its $j$th element which is one. Then, we can calculate the transition probabilities between SS as follows. First notice that the transition from $S_j$ to $S_i$ for $j\neq i$ occurs if the process $Y_{2,j}(t)$ is reached, and it is absorbed by $S_j$, and it can be expressed as
\begin{align}
    p_{ji}(\tau_j)&=\left(1-F_{T_{1,j}}(\tau_j)\right) \left[-\bm{\beta}_2^{(j)}\bm{A}_2^{(j)} \bm{B}_2^{(j)}e^{(j)} \right] \nonumber
    \\&=-\bm{\beta}_1^{(j)} e^{\tau_j\bm{A}_1^{(j)}} \bm{A}_2^{(j)} \bm{B}_2^{(j)}e^{(j)} \label{eq:pji_app}
\end{align}
for $j\neq i$. In order to calculate the self transition probability for $S_j$, we should update it by considering the absorption probability during $Y_{1,j}(t)$ as
\begin{align}
    p_{jj}(\tau_j)&=-\bm{\beta}_1^{(j)} e^{\tau_j\bm{A}_1^{(j)}} \bm{A}_2^{(j)} \bm{B}_2^{(j)}e^{(j)}+F_{T_{1,j}}(\tau_j) \nonumber \\
    =&-\bm{\beta}_1^{(j)} e^{\tau_j\bm{A}_1^{(j)}} \bm{A}_2^{(j)} \bm{B}_2^{(j)}e^{(j)}+1-\bm{\beta}_2^{(j)} e^{\tau_j \bm{A}_1^{(j)}}\bm{1}. \label{eq:pjj_app}
\end{align}

The absorbing CTMC whose generator is given in \eqref{eq:Tmatrix} can be converted to a DTMC with probability transition matrix as, 
\begin{align}
 \left[ \begin{array}{c|c}
   \bm D & \bm E \\
   \midrule
   \bm{0} & \bm{0} \\ 
\end{array}\right] \label{eq:TmatrixDiscrete1}
\end{align}
at the embedded epochs of state transitions. In particular, $\bm D = \{ d_{ij} \}$ can be written as, 
\begin{align}
    d_{ij}=\begin{cases}
        -\dfrac{a_{ij}}{a_{jj}}, & \text{if } i\neq j, \\ 0, & \text{otherwise.}
    \end{cases} \label{eq:diag}
\end{align}
A basic property about an absorbing Markov chain is that the expected number of visits to a transient state $j$ starting from a transient state $i$ is given by the $(i,j)$th entry of the so-called fundamental matrix $\bm F=(\bm I-\bm D)^{-1}$ \cite{kemeny1960finite}, where $\bm I$ denotes the identity matrix of suitable size. Therefore, the number of visit of transient state $i$ on $Y_{2,j}$ is calculated with
\begin{align}
    c_j(\tau)&=Pr(T_1>\tau)\bm{\beta}_2^{(j)} (\bm{I}-\bm D^{(j)})^{-1} \bm{1}\\
    &=\bm{\beta}_1^{(j)} e^{\bm{A}_1^{(j)}\tau_j} (\bm{I}-\bm D^{(j)})^{-1} \bm{1}, \label{eq:rj}
\end{align}
where $\bm{D}^{(j)}$ obtained from $\bm{A}_2^{(j)}$ as in Eqn.~\eqref{eq:diag}.

Finally, we conclude the section with examples of characteristic matrices AMCs $Y_{k,j}(t) \sim AMC(\bm {A_k}^{(j)},\bm{B_k}^{(j)},\bm{\beta_k}^{(j)})$ for $j=1$ obtained by $3 \times 3$ generator matrix $\bm{Q}$ with diagonal elements $q_{ii}=-\sigma_i$ as
\begin{align}
    {A_1}^{(1)}&=\begin{bmatrix}
        -\sigma_2 & q_{23} \\
        q_{32} & -\sigma_3
    \end{bmatrix}, \ {B_1}^{(1)}=\begin{bmatrix}
        q_{21} \\
        q_{31}
    \end{bmatrix} \\ 
    {A_2}^{(1)}&=\begin{bmatrix}
        -\sigma_2-\mu & q_{23} \\
        q_{32} & -\sigma_3-\mu
    \end{bmatrix}, \ {B_2}^{(1)}=\begin{bmatrix}
        q_{21} & \mu & 0\\
        q_{31} & 0 & \mu
    \end{bmatrix}
\end{align}

\begin{algorithm}
\caption{Lagrangian finding with bisection}\label{alg:lag}
\begin{algorithmic}
\State Apply Algorithm \ref{alg:cap} in the manuscript for $\lambda=0$ and obtain $\bm{\tau}$
\If{$R(\bm{\tau})\leq b$}
    \State $\bm{\tau}$ is optimum for constrained problem.
\Else
    \State $\lambda_{\text{mid}}\gets \frac{\lambda_{\text{max}}+\lambda_{\text{min}}}{2}$
    \While{$|R(\bm{\tau})-b|>\epsilon_{\lambda}$}
    \State Apply Algorithm \ref{alg:cap} in the manuscript for $\lambda_{\text{mid}}$ and obtain $\bm{\tau}$
    \If{$R(\bm{\tau})\geq b$}
    \State $\lambda_{\text{min}}\gets\lambda_{\text{mid}}$
    \Else
    \State $\lambda_{\text{max}}\gets\lambda_{\text{mid}}$
    \EndIf
    \State $\lambda_{\text{mid}}\gets\frac{\lambda_{\text{max}}+\lambda_{\text{min}}}{2}$
    \EndWhile
\EndIf
\end{algorithmic}
\end{algorithm}

\section{Finding the optimum Lagrangian coefficient} \label{app:lag}
The bisection algorithm that finds the optimum Lagrangian coefficient is summarized in Algorithm~\ref{alg:lag}. In each iteration, the unconstrained Lagrangian problem in \eqref{OptUc} is solved for a Lagrangian coefficient by using the Algorithm~ \ref{alg:cap} in the manuscript, then it is checked that if obtained thresholds satisfy the sampling rate constraints or not. For this purpose, we can express the values of  MAoII and $R$ for given threshold values $\bm{\tau}=\{\tau_j\}$ by following the same approach in \cite{cosandal2024modeling}, as
\begin{align}
   \mbox{MAoII}(\bm{\tau}) & =\dfrac{\sum_{j=1}^{N}\pi_j(\bm{\tau}) a_j(\tau_j)}{\sum_{j=1}^{N} \pi_j(\bm{\tau}) d_j(\tau_j)}, \label{eq:age_app} \\
     R(\bm{\tau})& =\dfrac{\sum_{j=1}^{N}\pi_j(\bm{\tau}) c_j(\tau_j)}{\sum_{j=1}^{N} \pi_j(\bm{\tau}) d_j(\tau_j)}, \label{eq:rate_app}
\end{align}
where $\pi_j(\bm{\tau})$ is the steady-state probability of being in SS $S_j$ at a SP given the vector of thresholds $\bm{\tau}$. Notice that, $\bm{\pi}(\bm{\tau})= \{ \pi_j(\bm{\tau})\}$ satisfies
\begin{align}
    \bm{\pi}(\bm{\tau})&=\bm{\pi}(\bm{\tau}) \bm{P}(\bm{\tau}), \quad \bm{P}(\bm{\tau}) \bm{1}=\bm{1}, 
\end{align}
where $\bm P(\bm{\tau})=\{ p_{ij}(\tau_j) \}$, they are obtained by Eqns. \eqref{eq:pji}-\eqref{eq:pjj} in the manuscript. Then, we obtain $\bm{\pi}(\bm{\tau})$ as,
\begin{align}
    \bm{\pi }(\bm{\tau}) &=\bm{1}^T(\bm{P}(\bm{\tau})+\bm{1}\bm{1}^T-\bm{I})^{-1}. \label{eq: oneonetranspose}
\end{align}

\end{document}